# Inferring protein-protein interaction and protein-DNA interaction directions based on cause–effect pairs in undirected and mixed networks[1]

Mehdy Roayaei[2] and MohammadReza Razzazi[3]

Department of Computer Engineering and Information Technology,

AmirKabir University of Technology,

Tehran, Iran

**Abstract**

We consider the following problem: Given an undirected (mixed) network and a set of ordered source-target, or cause-effect pairs, direct all edges so as to maximize the number of pairs that admit a directed source-target path. This is called maximum graph orientation problem, and has applications in understanding interactions in protein-protein interaction networks and protein-DNA interaction networks. We have studied the problem on both undirected and mixed networks. In the undirected case, we determine the parameterized complexity of the problem (for non-fixed and fixed paths) with respect to the number of satisfied pairs, which has been an open problem. Also, we present an exact algorithm which outperforms the previous algorithms on trees with bounded number of leaves. In addition, we present a parameterized-approximation algorithm with respect to a parameter named the number of backbones of a tree. In the mixed case, we present polynomial-time algorithms for the problem on paths and cycles, and an FPT-algorithm based on the combined parameter the number of arcs and the number of pairs on general graphs.

Keywords: protein-protein interaction, protein-DNA interaction, cause–effect pairs, fixed-parameter tractable, W[1]-hardness

---

[1] This research was in part supported by a grant from I.P.M.
[2] Corresponding author (mroayaei@aut.ac.ir). Tel: +982164542732, P.O. Box 15875-4413, #424 Hafez Avenue, AmirKabir University of Technology, Tehran, Iran
[3] razzazi@aut.ac.ir

# 1 Introduction

Protein-protein interactions (PPIs) form the skeleton of signal transduction in the cell. These interactions carry directed signaling information. However, current technologies [8], [11] cannot decide the direction in which the signal flows. Inferring the directions of these interactions is fundamental to our understanding of how these networks work. Perturbation experiments [16] provide additional information for possible direction of information in these networks. In this experiment, a gene is perturbed (cause) and as a result other genes change their expression levels (effects), to guide the orientation inference. It is assumed that there must be a directed path in the network from the causal gene (source) to the affected gene (target). The resulting combinatorial problem, which is called the maximum graph orientation, is to orient the edges of the network such that a maximum number of cause-effect pairs admit a directed path from the causal to the affected gene. When studying a PPI network in isolation, the input network is undirected. However, the more biologically relevant variant considers also protein-DNA interactions as these are necessary to explain the expression changes. Moreover, the directionality of some PPIs is known in advance [6]. Therefore, generally, the input network is considered as a mixed graph containing both directed and undirected edges. In this paper, we consider the maximum graph orientation on both undirected and mixed graphs.

The maximum graph orientation problem on undirected graphs and mixed graphs is defined as follows:

**Definition 1 (Maximum undirected graph orientation problem- MUGO).** An undirected graph $G = (V, E)$, and a set $P = \{(s_i, t_i): 1 \leq i \leq p\}$ of ordered source-target pairs are given, where $E$ is the set of edges. Direct all edges so as to maximize the number of pairs that admit a directed source-target path.

**Definition 2 (Maximum mixed graph orientation problem- MMGO).** A mixed graph $G = (V, E, A)$, and a set $P = \{(s_i, t_i): 1 \leq i \leq p\}$ of ordered source-target pairs are given, where $E$ is the set of edges and $A$ is the set of arcs. Direct all edges so as to maximize the number of pairs that admit a directed source-target path.

In the remainder of the paper, let $P$ be the set of input pairs, $|P| = p$, $V$ the set of vertices of a tree or a graph, and $|V| = n$. By edge, we mean an undirected edge, and by arc, we mean a directed edge. A pair is *satisfied*, if it admits a directed source-target path.

It is shown than an MUGO problem can be converted to an equivalent problem on a tree, which is obtained by contracting cycles of the input graph [12]. Thus, the interesting case is when the input graph is a tree. This problem is called *maximum tree orientation* (MTO). MTO is NP-hard, even on stars, caterpillars, and binary trees [12], but it is polynomial-solvable on paths. The best approximation ratio obtained for MTO is $\Omega(\frac{\log \log n}{\log n})$ [10]. It is NP-hard to approximate MTO within a factor of $\frac{11}{12}$. Also, MTO has been studied from the parameterized complexity point of view. It is shown that this problem is fixed-parameter tractable with respect to the parameters the maximum number of pairs passing through a vertex, and the maximum number of cross pairs passing through a vertex (the cross pair is defined as a source-target pair whose corresponding path is directed either towards the root or towards the leaves, but do not change its direction.) [4]. However, it is not fixed-parameter tractable w.r.t. the maximum number of pairs passing through an edge.

MMGO is also NP-hard. Furthermore, although MTO, the feasibility version of MMGO, the problem of deciding whether a mixed graph *G* can be oriented so that the resulting directed graph contains a directed source-target path for all input pairs, is NP-complete [1]. The best approximation ratio obtained for MMGO is $\Omega(\frac{1}{(n|P|)^{1/3}})$ [9]. It is NP-hard to approximate MMGO within a factor of $\frac{7}{8}$ [6]. To our knowledge, there has been no significant results on the parameterized complexity of MMGO problem.

An *orientation* of a graph is an assignment of a direction to each edge. We say that the pair $(s_j, t_j)$ conflicts with the pair $(s_i, t_i)$ if there exist no orientation of the input graph for which both $(s_j, t_j)$ and $(s_i, t_i)$ are satisfied.

We call an tree with exactly one vertex of degree more than two, a *star-like tree*. *Split vertex* is a vertex of degree two, whose incident arcs have different directions. After orienting an undirected path between the vertices *v* and *w*, the nearest split vertex to the vertex *v* on that path, is denoted by $split(v, w)$.

A path between two vertices *a* and *b* is denoted by a-to-b or [a-to-b]. If this path does not contain the vertex *a* (*b*) and its incident edge, it is denoted by (a-to-b] ([a-to-b)). A path between a vertex *v* of degree more than two and a vertex of degree one, is called a *branch* incident to the vertex *v*. A vertex *v* is *far-adjacent* to a vertex *w*, if by ignoring the vertices of degree two, the vertex *v* is adjacent to the vertex *w*. Let $T_1$ and $T_2$ be two subtrees of the tree *T*. The graph resulting from the merging of these two subtrees is denoted by $T_1 + T_2$. A graph is called $K_4$-free, if it does not contain any clique of size 4.

Parameterized computation is a new approach dealing with NP-hard problem [2], [17], and [3]. A fixed-parameter tractable algorithm (FPT-algorithm) is an algorithm that solves a problem of input size *n* and a parameter *k* in $f(k).n^{O(1)}$ time, in which *f* is a computable function depending only on the parameter *k* [13]. If a problem is W[1]-hard with respect to a parameter *k*, then it means there is no FPT-algorithm for it (unless $FPT = W[1]$).

The remainder of the paper is organized as follows. In section 2, the MUGO problem on undirected graphs is studied. We determine the parameterized complexity of the problem (for non-fixed and fixed paths) w.r.t. the number of satisfied pairs, which has been an open problem. Also, we present an exact algorithm which outperforms the previous algorithms on trees with a limited number of leaves. In addition, we present a parameterized-approximation algorithm w.r.t. a parameter named the number of backbones of a tree. In section 3, the MMGO on mixed graphs problem is studied. We present polynomial-time algorithms for paths and cycles, and an FPT-algorithm based on the combined parameter the number of arcs and the number of pairs for general graphs. We conclude our paper in section 4 by introducing some open problems.

## 2 Undirected Networks

In this section, we study the complexity of the MUGO problem w.r.t. the number of satisfied pairs, the number of leaves of the input tree, and a parameter named the number of backbones of the input tree.

### 2.1 Number of satisfied pairs

The parameterized complexity of MUGO problem w.r.t. the number of satisfied pairs has been an open problem [4]. In this subsection, we determine the parameterized complexity of this problem for fixed and non-fixed paths. The fixed-path variant of MUGO is identical to MUGO with the exception that each pair $(s_j, t_j) \in P$ is associated with a fixed path $p_j$ from $s_j$ to $t_j$ in the graph. Hence, a pair $(s_j, t_j)$ is satisfied only if the edges of the path $p_j$ is oriented from the vertex $s_j$ towards the vertex $t_j$.

First, we study the problem for non-fixed paths. Since paths are non-fixed, we can assume that the input graph is a tree.

Each instance of the MTO problem can be modeled as an instance of the *Maximum Independent Set* (MIS) problem. An independent set is a set of vertices in a graph, no two of which are adjacent. A maximum independent set is an independent set of largest possible size for a given graph. Each pair $(s_j, t_j)$ of MTO instance is considered as a vertex $v_j$ in the MIS instance. There is an edge between two vertices of MIS instance, if and only if the corresponding pairs conflict with each other. We call the resulting graph a *conflict graph*. It is clear that finding the optimal solution of the MIS problem on the conflict graph is equivalent to finding the optimal solution of the corresponding MTO instance.

**Lemma 2.1.1** [4]**.** The resulting conflict graph of an MTO instance is $K_4$-free.

Now, we model the MIS problem as the *Party problem* [15]: find the minimum number of guests that must be invited so that at least $\alpha$ guests will know each other (a clique of size $\alpha$) or at least $\beta$ guests will not know each other (an independent set of size $\beta$). The solution of the problem is known as the ramsey number $R(\alpha, \beta)$. Thus, if we consider each vertex of the conflict graph as a guest, and each edge between two vertices as the corresponding guests knowing each other, the rumsey number $n = R(\alpha, \beta)$ returns the minimum number of vertices such that a graph with at least $n$ vertices contains a clique of size $\alpha$ or an independent set of size $\beta$. Ramsey's theorem states that such a number exists for all $\alpha$ and $\beta$.

**Lemma 2.1.2** [7]. $R(\alpha, \beta) \leq \binom{\alpha+\beta-2}{\beta-1}$.

According to the lemma 2.1.1, the conflict graph whose the number of vertices is equal to or more than $R(4, \beta)$, has an independent set of size at least $\beta$. According to lemma 2.1.2, $R(4, \beta) \leq \frac{(\beta+2)(\beta+1)(\beta)}{6}$. Hence, a conflict graph for which $n \geq \frac{(\beta+2)(\beta+1)(\beta)}{6}$, has an independent set of size at least $\beta$.

**Theorem 2.1.1.** The MTO problem is fixed-parameter tractable with respect to the maximum number of satisfied pairs.

*Proof.* Let $\beta$ be the maximum number of satisfied pairs. First, create the conflict graph $T_c$ of the MTO problem. If $p > \frac{(\beta+2)(\beta+1)(\beta)}{6}$, we can remove arbitrary $p - \frac{(\beta+2)(\beta+1)(\beta)}{6}$ vertices from $T_c$. According to Ramsey's theorem, $T_c$ has still an independent set of size at least $\beta$. Thus, regardless of the size of $T_c$, it can be reduced to a kernel of size at most $\frac{(\beta+2)(\beta+1)(\beta)}{6}$ such that there is an independent set of size $\beta$ in the kernel, if and only if there is an independent set of size $\beta$ in $T_c$. Since we have reduced the input instance to a polynomial-size kernel w.r.t. to the parameter $\beta$, MTO is fixed-parameter tractable w.r.t. the maximum number of satisfied pairs. □

We show that, despite the non-fixed version, the MUGO problem with fixed paths is W[1]-hard w.r.t. the number of satisfied pairs. We reduce from the *K-clique* problem: Given an undirected graph $G_c = (V_c, E_c)$, and a parameter $K$, is there a clique (a set of vertices that are pairwise adjacent) of size $K$ in $G_c$? The K-clique problem is W[1]-hard w.r.t. the parameter $K$ [5].

Given an instance $G_c = (V_c, E_c)$ of *K*-clique problem, we construct an instance $G = (V, E)$ of MUGO as follows. For each vertex $v_i \in V_c$, we create two vertices $s_i$ and $t_i$ such that the pair $(s_i, t_i)$ is connected by a fixed path $p_i$. If there is no edge between the two vertices $v_i, v_j \in V_c$, intersect the corresponding two paths $p_i$ and $p_j$ as shown in Figure 1 (a). In this case, the path $p_i$ contains the edge $u_{ij}$-$v_{ij}$, and the path $p_j$ contains the subpath $u_{ji}$-$v_{ij}$-$u_{ij}$-$v_{ji}$. Otherwise, if there is an edge between two vertices $v_i, v_j \in V_c$, pass one of them above the other, as shown in Figure 1 (b). In this case, the path $p_i$ contains the edge $u_{ij}$-$v_{ij}$, and the path $p_j$ contains the edge $u_{ji}$-$v_{ji}$.

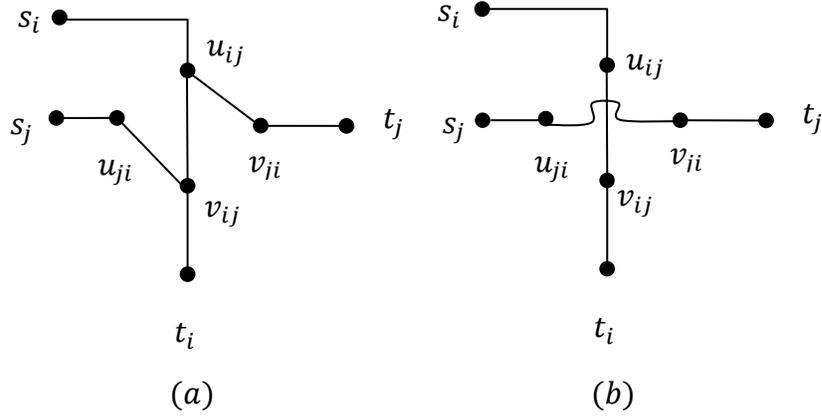

**Figure 1. Intersection of two paths in an MUGO instance**

In Figure 2, a graph of size five, which contains a 3-clique, and its corresponding MUGO instance is depicted. One can see that the three vertices $v_2$, $v_3$, and $v_5$ of the 3-clique in $K$-clique instance are one-to-one corresponding to the three paths $p_2$, $p_3$, and $p_5$ that can be satisfied at the same time in the MUGO instance. The following theorem is the conclusion of the discussion above.

**Theorem 2.1.2.** The maximum graph orientation problem with fixed paths is W[1]-hard w.r.t. the number of satisfied pairs.

*Proof.* According to the discussion above, we reduce from the $K$-clique problem. Since the $K$-clique problem is W[1]-hard w.r.t. the parameter $K$, and the number of satisfied pairs in the MUGO instance is equal to the size of a clique in the $K$-clique instance, the MUGO problem with fixed paths is W[1]-hard w.r.t. the number of satisfied pairs. □

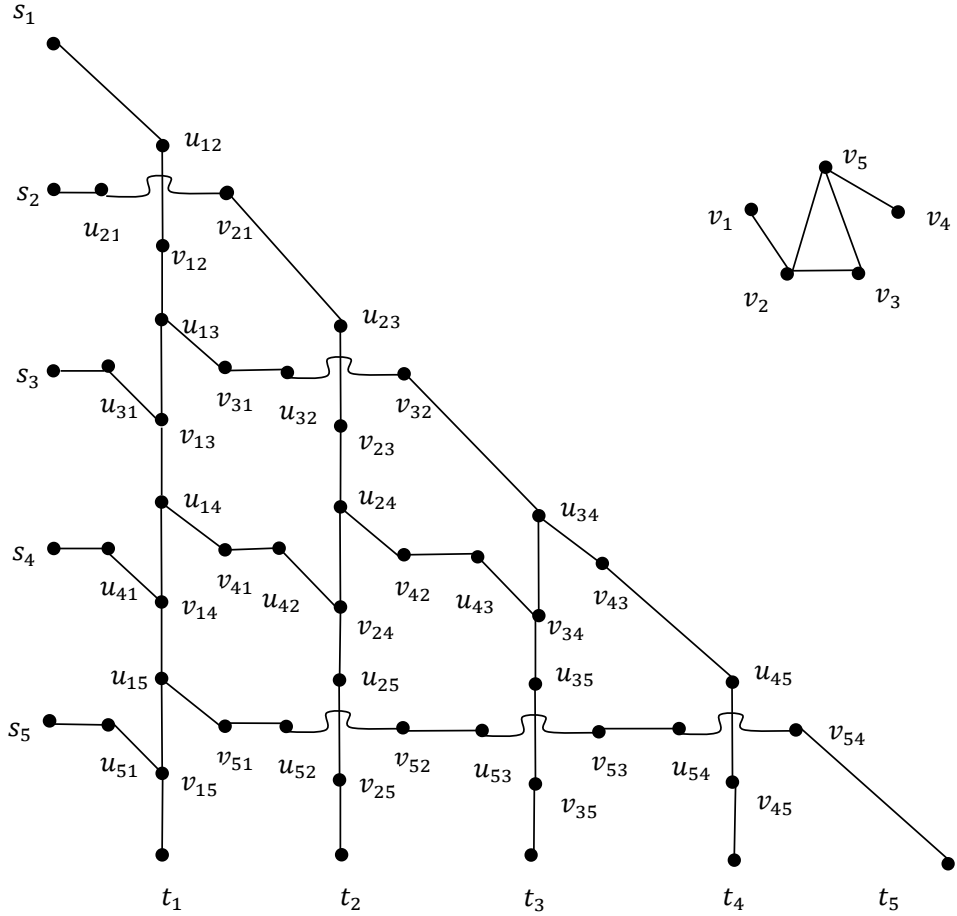

**Figure 2. An instance of clique problem and its corresponding MUGO instance**

## 2.2 Number of leaves

In [4], the parameterized complexity of MTO was studied w.r.t. to the maximum signal flow over vertices or edges. They defined the notion of *cross pair* as a source-target pair whose corresponding path is directed either towards the root or towards the leaves, but do not change its direction. They showed that MTO is fixed-parameter tractable with respect to the maximum number of cross pairs passing through a vertex, denoted by $q_v$, and presented an $O(2^{q_v} q_v n^2)$ time algorithm to solve it.

In this subsection, we present an exact algorithm which outperforms the mentioned algorithm on the trees with bounded number of leaves. In fact, We show that the MTO problem is tractable for a constant number of leaves.

We use the following straightforward lemmas in this subsection.

**Lemma 2.2.1.** Maximum tree orientation can be solved on paths in $O(n^2)$ time [4].

**Lemma 2.2.2.** The number of vertices of degree more than two in a tree is at most the number of leaves minus two.

Consider the star-like tree in Figure 3 (a), which has three leaves.

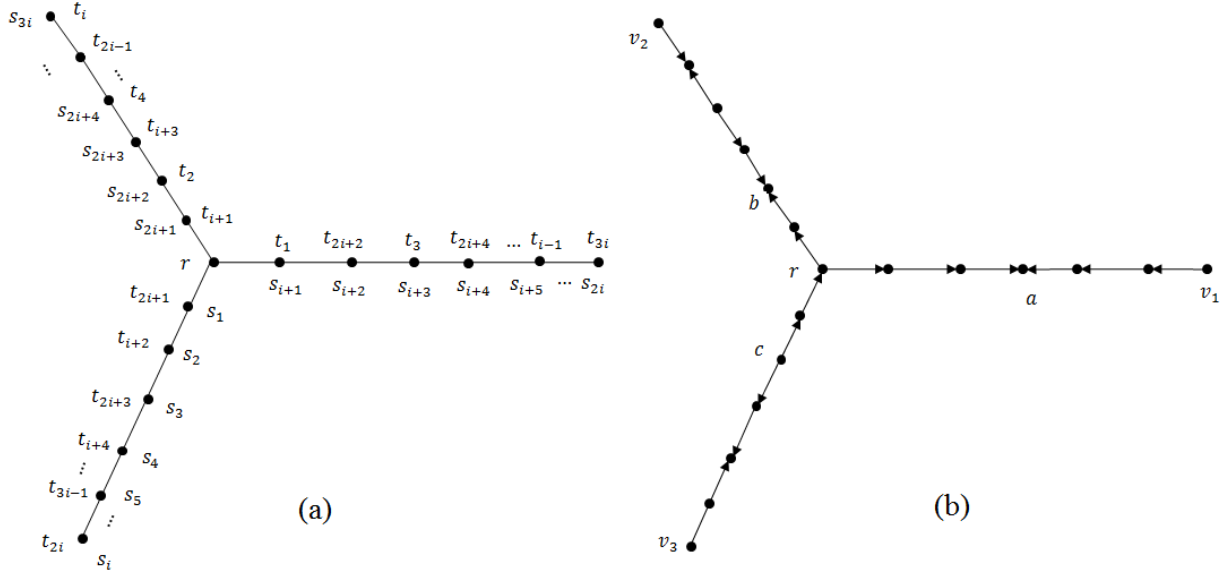

**Figure 3. star-like trees with three leaves**

Without loss of generality, assume that $n = 3i + 1$, for some even integer $i$; and there are $i$ vertices on each branch of the tree. Regardless of the vertex selected as the root, the maximum number of cross pairs passing through a vertex is $O(n)$. Hence, using the FPT-algorithm based on the parameter $q_v$, this instance is solved in $O(2^n)$ time. We present another algorithm that solves this instance in $O(n^5)$.

We illustrate the core idea of our algorithm in Figure 3 (b). Consider the path between the vertex $r$ and the leaf $v_1$. In an arbitrary orientation of the tree, when we walk away from the vertex $r$ to the vertex $v_1$, all arcs are in the same direction until we reach to an arc with the different direction (that is, until a split vertex appears). If there is not a split vertex in the path $r$-to-$v_1$, it means that all arcs on this path have the same direction. Now, as in Figure 3 (b), assume that the first split vertex on the path (from $r$-to-$v_1$) is the vertex $a$. One can see that there cannot be a satisfied pair $(s_j, t_j)$ such that $s_j \in$ path $(a$-to-$v_1]$ and $t_j \notin$ path $[a$-to-$v_1]$ or vice versa (that is, $t_j \in$ path $(a$-to-$v_1]$ and $s_j \notin$ path $[a$-to-$v_1]$). Hence, we can process the subpath $a$-to-$v_1$ independently.

This approach can be applied on all branches of the tree. For each branch, there are at most $O(n)$ possible choices for choosing the first split vertex on that branch. If there is no split vertex on a branch, the leaf of the branch is considered as its first split vertex. Hence, there are at most $O(n^3)$ possible combinations for choosing the first split vertices of the branches of the tree. For each combination, the tree is decomposed into three independent paths ($a$-to-$v_1$, $b$-to-$v_2$, and $c$-to-$v_3$), called *separated paths*, and a star-like tree with the leaves $a$, $b$, and $c$, where all edges of each branch have the same direction. The branches $r$-to-$a$, $r$-to-$b$, and $r$-to-$c$ are called *new branches* of the tree. According to lemma 2.2.1, the separated paths can be processed simultaneously in $O(n^2)$ time. Also, the newly generated star-like tree can be processed in polynomial time. Therefore, the MTO problem on a tree with three leaves can be solved in $O(n^5)$ time. This algorithm can be easily generalized for all star-like trees with $k$ leaves, in $O(n^{k+2}.2^k.k)$ time.

We show how our algorithm can be generalized for all trees. Consider the situation, where there is more than one vertex of degree more than two, as in Figure 4(a). First, consider the vertex $r_1$. Assume that the vertices $a$, $b$, and $c$ are the split vertices of the branches incident to $r_1$. As in the previous case, we decompose the tree at these vertices, and orient each of the new branches ($r_1 - a$, $r_1 - b$, and $r_1 - c$). Note that for each separated path, the set of input pairs contains the pairs for which both endpoints are on that path. Also, the remaining subtree can be processed independently. However, the set of input pairs $P$ must be updated for this subtree. For each pair $(s_j, t_j) \in P$, one of the following cases occurs:

- Pairs such as $(s_j, t_j)$, for which $s_j$ and $t_j$ are on one of the separated subpaths. Such pairs must be deleted from the set $P$.
- Pairs such as $(s_j, t_j)$, for which $s_j$ (or $t_j$) belongs to one of the separated subpaths (as $(s_1, t_1)$ in Figure 4 (a)). Such pairs cannot be satisfied, thus, are removed form $P$.
- Pairs such as $(s_j, t_j)$, for which $s_j$ (or $t_j$) belongs to one of the new branches, but cannot be satisfied according to the current orientation (as $(s_2, t_2)$ in Figure 4(a)). Such pairs are also removed from $P$.
- Pairs such as $(s_j, t_j)$, for which $s_j$ (or $t_j$) belongs to one of the new branches, and may be satisfied according to the current orientation. (as $(s_3, t_3)$ in Figure 4 (a)). In this case, $s_j$ (or $t_j$) is transformed to the corresponding vertex of degree more than two (as $s_3$ is transformed to $r_1$ in Figure 4 (b)). In this case, the pair $(s_j, t_j)$ in $P$ is replaced by the new pair.
- Other pairs, which are remained in $P$.

After updating the set $P$ of source-target pairs for each combination of split vertices and new branches orientation, the new branches are removed from the subtree obtained from the decomposition of the tree, as in Figure 4 (b).

Thus, after selecting the split vertices of the branches of $r_1$, decomposing the tree at the split vertices, and then orienting its new branches, the remaining subtree, which has one fewer vertex of degree more than two, can be processed independently. Note that there is always a vertex of degree more than two, which is far-adjacent to exactly one vertex of degree more than two. We start the processing of the tree by choosing such vertex (as $r_1$ in Figure 4 (a)). Then, this process is iteratively performed until the whole tree is processed.

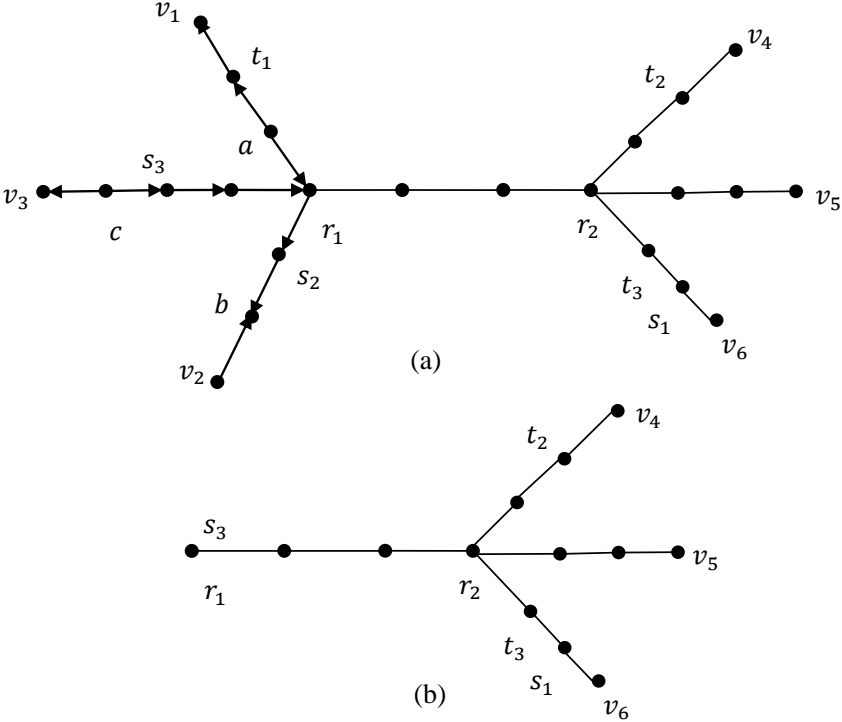

**Figure 4. Illustration of the algorithm on general graphs**

**Theorem 2.2.1**. Maximum tree orientation can be solved on trees in $O((2n)^{2k})$ time, where $k$ is the number of leaves of the tree.

*Proof.* As described above, in each iteration of the algorithm, we choose a vertex of degree more than two that is far-incident to exactly one vertex of degree more than two. We determine the split vertices on its branches, decompose the tree at these vertices, orient the new branches, update the source-target pairs, and remove the new branches. This process continues until the whole tree is processed.

The number of split vertices that are selected for branches of the tree during the algorithm, are corresponding to the number vertices of degree one or degree more than two. According to lemma 2.2.2, the total number of these split vertices is at most $2k - 2$. Thus, there are $O(n^{2k-2})$ combinations for selecting the split vertices. Also, there are two possible orientations for each new branch which is obtained after the decomposition of the tree at a split vertex. Therefore, there are $O(2^{2k-2})$ combinations for selecting the orientations of new branches. Altogether, the algorithm runs in $O(n^{2k} \cdot 2^{2k-2})$ time. Note that we have omitted some polynomial factors from the time complexity of the algorithm to simplify it. This algorithm is depicted in Figure 5. □

**Algorithm 1:** MTO-LEAVES

**Input**: An undirected tree $T = (V, E)$ and a set $P = \{(s_i, t_i) : 1 \leq i \leq p\}$ of source-target pairs
**Output**: direct all edges to maximize the number of satisfied pairs

1 **if** *there is more than one vertex of degree more than two in $T$* **then**
2     select a vertex $r$ of degree $> 2$ such that it is far-adjacent to exactly one vertex of degree $> 2$
3     $max = 0$; $T_{max} = \emptyset$; $i = 0$;
4     **for** *each combination of split vertices of the branches incident to $r$ in $T$* **do**
5        $i = i + 1$;
6        $T_{sol_i} = \emptyset$ ; $P_{sat_i} = \emptyset$ ; $T_i = T$; $P_i = P$;
7        **for** *each branch $r$-to-$v$* **do**
8           decompose $T_i$ at $split(r, v)$, and optimally orient the subpath $split(r, v)$-to-$v$;
9           add the oriented subpath to $T_{sol_i}$, and the satified pairs to $P_{sat_i}$;
10           update $P_i$;
11        $j = 0$ ; $max_i = 0$ ; $T_{max_i} = \emptyset$;
12        **for** *each orientation of the new branches incident to $r$* **do**
13           $j = j + 1$ ;
14           $T_{sol_{i,j}} = T_{sol_i}$ ; $P_{sat_{i,j}} = P_{sat_i}$ ; $T_{i,j} = T_i$ ; $P_{i,j} = P_i$ ;
15           update $P_{i,j}$ according to the orientation of new branches incident to $r$ ;
16           remove new branches incident to $r$ from $T_{i,j}$ and add them to $T_{sol_{i,j}}$;
17           **if** *number of satified pairs in $MTO - LEAVES(T_{i,j}, P_{i,j}) + |P_{sat_{i,j}}| > max_i$* **then**
18              $max_i$ = number of satified pairs in $MTO - LEAVES(T_{i,j}, P_{i,j}) + |P_{sat_{i,j}}|$ ;
19              $T_{max_i} = T_{sol_{i,j}} + MTO - LEAVES(T_{i,j}, P_{i,j})$ ;
20        **if** $max_i > max$ **then**
21           $max = max_i$ ; $T_{max} = T_{max_i}$ ;
22     return $T_{max}$;
23 **else if** *there is exactly one vertex of degree more than two in $T$* **then**
24     select the vertex $r$ of degree more than two in $T$;
25     $max = 0$ ; $T_{max} = \emptyset$ ; $i = 0$ ;
26     **for** *each combination of split vertices of the branches incident to $r$ in $T$* **do**
27        $i = i + 1$ ;
28        $T_{sol_i} = \emptyset$ ; $P_{sat_i} = \emptyset$ ;
29        **for** *each branch $r$-to-$v$* **do**
30           decompose $T$ at $split(r, v)$, and optimally orient the subpath $split(r, v)$-to-$v$ ;
31           add the oriented subpath to $T_{sol_i}$, and add the satified pairs to $P_{sat_i}$ ;
32        $j = 0$ ; $max_i = 0$ ; $T_{max_i} = \emptyset$;
33        **for** *each orientation of the new branches incident to $r$* **do**
34           $j = j + 1$ ;
35           $T_{sol_{i,j}} = T_{sol_i}$ ; $P_{sat_{i,j}} = P_{sat_i}$ ;
36           add the satified pairs to $P_{sat_{i,j}}$ ;
37           add the new branches to $T_{sol_{i,j}}$ ;
38           **if** $|P_{sat_{i,j}}| > max_i$ **then**
39              $max_i = |P_{sat_{i,j}}|$ ; $T_{max_i} = T_{sol_{i,j}}$ ;
40        **if** $max_i > max$ **then**
41           $max = max_i$ ; $T_{max} = T_{max_i}$ ;
42     return $T_{max}$ ;
43 **else**
44     optimally orient $T$ as $T_{max}$, and return $T_{max}$ ;

**Figure 5. The algorithm based on the number of leaves**

## 2.3 Number of backbones

As stated before, the best approximation ratio obtained for MTO is $\Omega(\frac{\log \log n}{\log n})$ [10], and there has been no constant-ratio approximation algorithm for it. In this section, we introduce the notion of the *backbone* in a tree, and present a parameterized-approximation algorithm with respect to the number of backbone of a tree, which provides a constant ratio for trees with bounded number of backbones.

A *caterpillar* is a tree in which all vertices are within distance one of a central path. We define a *caterpillar-like* tree as a tree in which all vertices of degree more than two are on the same path. We call this central path, a backbone.

Consider the star-like tree and the caterpillar-like tree in Figure 6.(a). In the star-like tree, orient each branch randomly. Since each pair passes through at most two branches, thus, each pair is satisfied with a probability of at least $1/4$. In the caterpillar-like tree, orient the central path (the backbone) $v_1$-to-$v_2$ and each of the branches incident to it randomly. Analogously, each pair is satisfied with a probability of at least $1/8$.

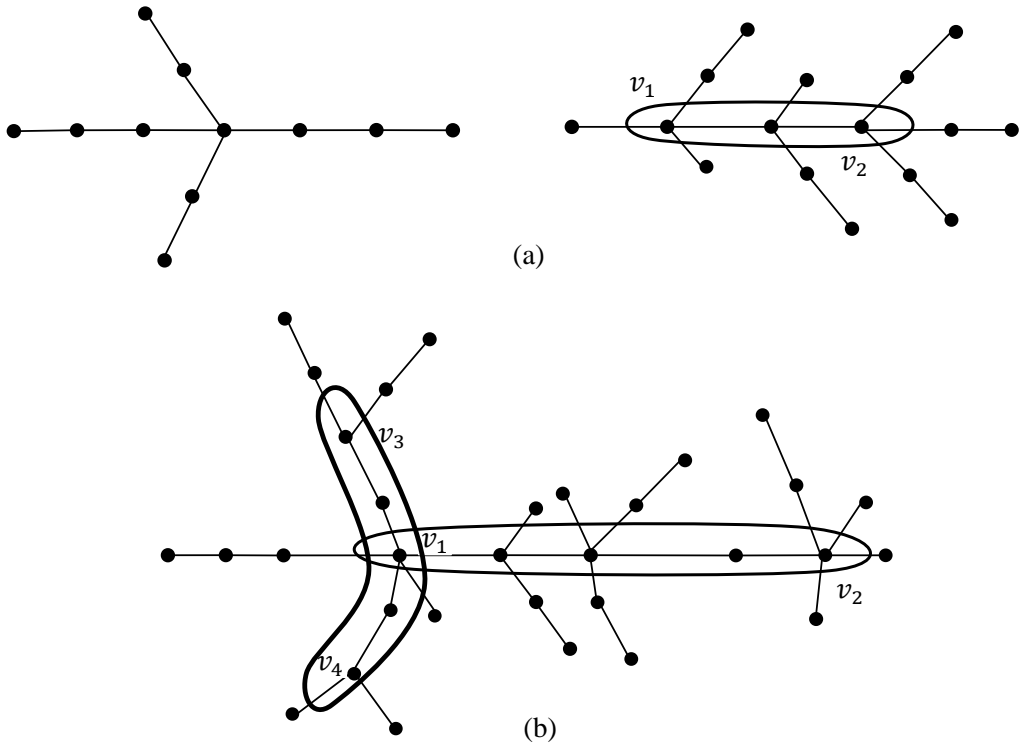

**Figure 6.** (a) a star-like and a caterpillar-like tree (b) a tree with two backbones

In general trees, vertices of degree more than two may be on more than one path (backbones). Note that these backbones must be edge-disjoint so that we can orient each of them randomly. For example, consider the tree in Figure 6 (b), in which, the vertices of degree more than two are at least on the two backbones $v_1$-to-$v_2$ and $v_3$-to-$v_4$. Assume that the vertices of degree more than two of a tree reside on $b$ edge-disjoint backbones. Then, by a random orientation of backbones and branches incident to the backbones, each pair is satisfied with a probability of at least $\frac{1}{2^{b+2}}$.

Now, we must find the minimum number of edge-disjoint backbones that cover all vertices of degree more than two. The first step is to contract all edges that are incident to a vertex of degree one or two. The resulting tree, which we call the backbone tree $T_b$, shows the far-adjacency of the vertices of degree more than two. It is obvious that a decomposition of $T_b$ into $b$ edge-disjoint paths provides $b$ edge-disjoint backbones that cover all vertices of degree more than two in $T$. So, it is sufficient to decompose $T_b$ into the minimum number of edge-disjoint paths.

**Lemma 2.3.1.** Let $T$ be a tree with $2b, b > 0$, vertices of odd degree. Then, $T$ can be decomposed into $b$ edge-disjoint paths. Also, any decomposition of $T$ into edge-disjoint paths contains at least $b$ paths.

*Proof.* The proof is by induction on $b$. If $b = 0$, $T$ has no edge. Assume that a tree with $2b - 2$ vertices of odd degree can be decomposed into $b - 1$ paths. Let a tree $T$ be a tree with $2b$ vertices of odd degree. $T$ has at least two vertices of degree one, namely $v_1$ and $v_2$. Deleting the edges of the path $v_1$-to-$v_2$, does not change the parity of the vertices except $v_1$ and $v_2$. Let the resulting tree be $T_{b-1}$. $T_{b-1}$ is a tree with $2b - 2$ vertices of odd degree, which can be decomposed into $b - 1$ paths. Thus, using the path $v_1$-to-$v_2$, $T$ can be decomposed into $b$ edges-disjoint paths.

Since adding an edge-disjoint path to a tree will increase the number of vertices of odd degree at most by 2, any decomposition contains at least $b$ edge-disjoint paths; otherwise, the resulting tree has less than $2b$ vertices of odd degree. □

The algorithm is illustrated in Figure 7.

```
Algorithm 2: MTO-BACKBONE-RANDOM
  Input: An undirected tree T = (V, E), and a set P = {(s_i, t_i) : 1 ≤ i ≤ p} of source-target
         pairs
  Output: An orientation for T to satisfy at least 1/2^{b+2} of pairs of P, where b is the minimum
          number of backbones of T that covers all vertices of degree more than two
1 Contract all edges incident to the vertices of degree one or two to obtain T_b;
2 Decompose T_b into the minimum number, b, of edge-disjoint paths;
3 for each edge-disjoint path of T_b do
4   | Orient randomly the corresponding backbone in T;
5 return the oriented T
```

**Figure 7. The algorithm based on the number of backbones**

It is easy to use the method of conditional expectations to obtain a deterministic algorithm from the algorithm in Figure 7 that produces an orientation for a graph with $b$ backbones that satisfies at least $\frac{1}{2^{b+2}}$ of the pairs.

**Theorem 2.3.1.** There is an approximation algorithm with the approximation ratio $\frac{1}{2^{b+2}}$ for the MTO problem, where $b$ is the minimum number of edge-disjoint backbones (paths) that cover all vertices of degree more than two.

## 3 Mixed Networks

Unlike the MTO problem, there may be more than one path between the source and the target of a pair of $P$ in an MMGO instance. Thus, there has been no efficient algorithm for determining conflicts between the pairs in $P$. Therefore, we state the following conjecture:

**Conjecture 3.1.** MMGO is not fixed-parameter tractable w.r.t. the number of pairs.

When the input is restricted to trees, because each pair corresponds to a path, MMGO is fixed-parameter tractable w.r.t. the number of pairs.

**Proposition 3.1.** MMGO on trees is fixed-parameter tractable w.r.t. the number of pairs.

### 3.1 Paths

Assume that a path is considered from left to right, and the vertices are numbered from 1 to $n$. For all $v, w \in V$, where $v \leq w$, $S(v, w)$ is the maximum number of pairs with both endpoints on the path $v$-to-$n$ that can be satisfied on the path $v$-to-$n$ such that the subpath $v$-to-$w$ is oriented from $v$ to $w$. Analogously, $S(w, v)$ is the maximum number of pairs with both endpoints on the path $v$-to-$n$ that can be satisfied on the path $v$-to-$n$ such that the subpath $v$-to-$w$ is oriented from $w$ to $v$. Also, $A(v, w)$ is the number of pairs with both endpoints on the path $v$-to-$w$ that are satisfied when orienting the path from $v$ to $w$.

When there is an arc on the path $v$-to-$w$ whose direction is toward $v$, then $A(v,w) = S(v,w) = 0$. Analogously, when there is an arc on the path $v$-to-$w$ whose direction is toward $w$, then $A(w,v) = S(w,v) = 0$. Also, $A(v,v) = 0$ for all $v$.

Then, $S(v,w)$ and $S(w,v)$ can be calculated as follows:

$$S(v,w) = A(v,w) + \max\{S(u,w), S(v,u) - A(v,w)\} \quad (3.1.1),$$

$$S(w,v) = A(w,v) + \max\{S(w,u), S(u,v) - A(w,v)\} \quad (3.1.2),$$

where $u$ is the right-hand side vertex of $w$, that is $u = w + 1$. Note than when $w = n$, then $S(v,w) = A(v,w)$, and $S(w,v) = A(w,v)$.

The main idea behind the recurrence relations (3.1.1) and (3.1.2) is that when a split vertex appears on a path, the path can be decomposed at that vertex and the two resulting subpaths can be processed independently. The following theorem is straightforward.

**Theorem 3.1.1.** MMGO on paths can be solved in $O(n^2)$ time.

*Proof.* In the first step, initialization, as stated above, is done in $O(n^2)$ time. The matrix $A$ can be easily computed in $O(n^2)$ time. Also, the matrix $S$ is computed in $O(n^2)$ time. $S(1,1)$ returns the value of the optimal solution of the MMGO problem on the input path. □

## 3.2 Cycles

If a cycle has no arc, or has one arc, or all of its arcs have the same direction, it can be oriented such that all pairs are satisfied. In this case, it is sufficient to orient all edges such that all edges (and all arcs) have the same direction.

Consider the cycle in Figure 8, in which there are two arcs with opposite directions. Because of these two arcs, there must be at least two split vertices in any optimal orientation. There are $O(n^2)$ combinations for choosing these two split vertices. Assume that $a$ and $b$ are those two split vertices. We can decompose the cycle at the split vertices into two subpaths between $a$ and $b$. The two subpaths are independent of each other except for the pair $(a,b)$. The pair $(a,b)$ is satisfied if and only if the path is oriented from $a$ to $b$.

$$|ab| = \max \begin{cases} |ab_1{}^+| + |ab_2{}^+| - \alpha_{(a,b)}, |ab_1{}^+| + |ab_2|_{P-(a,b)}, \\ |ab_2{}^+| + |ab_1|_{P-(a,b)}, |ab_1|_{P-(a,b)} + |ab_2|_{P-(a,b)} \end{cases} \quad (3.2.1),$$

where $|ab_1|_{P-(a,b)}(|ab_2|_{P-(a,b)})$ is the value of the optimal solution of the MMGO on the path $ab_1(ab_2)$ with input pairs $P - (a,b)$. Also, $\alpha_{(a,b)} = 1$, if the pair $(a,b) \in P$, otherwise $\alpha_{(a,b)} = 0$.

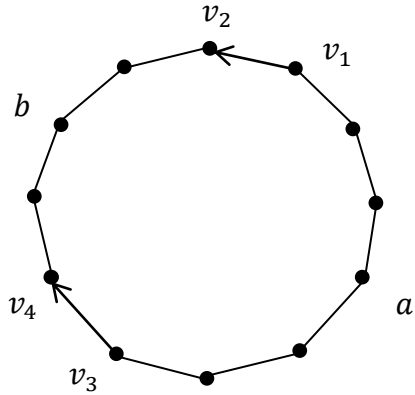

**Figure 8. Two arcs with opposite directions in a cycle**

This approach can be easily generalized for all cycles.

**Theorem 3.2.1.** Given a cycle $C = (V, E, A)$, and a set $P = \{(s_i, t_i): 1 \leq i \leq p\}$ of source-target pairs. The MMGO problem can be solved on $C$ in $O(n^4)$ time.

*Proof.* The algorithm is depicted in Figure 9. The algorithm is similar to the example showed above. The only difference is that a path $ab_j$ ($j = 1,2$) may have an arc with the direction opposite to the direction from *a* to *b*. In this case, we assign $|ab_j^+| = 0$.

There are $O(n^2)$ combinations for choosing the two split vertices. Since the two subpaths are independent of each other, for each combination, the solution can be calculated in $O(n^2)$ time. Thus, the MMGO problem on cycles can be solved in $O(n^4)$ time. □

```
Algorithm 3: MMGO-CYCLE
   Input: A mixed cycle C = (V, E, A), and a set P = {(s_i, t_i) : 1 ≤ i ≤ p} of source-target pairs
   Output: An orientation for C to satisfy maximum number of pairs
 1 if C has no arc, or has one arc, or all of its arcs have the same direction then
 2     orient all edges of C such that all edges (and all arcs) have the same direction;
 3     return the oriented C;
 4 else
 5     i = 0; max = 0;
 6     for each two split vertices a, b ∈ V do
 7         i = i + 1;
 8         max_i = 0; C_i = C;
 9         denote the two resulting paths by ab_1 and ab_2;
10         optimally orient the paths ab_1 and ab_2 for the set of input pairs P − (a, b) using
           theorem 3.1.1;
11         for the path ab_j, j = 1, 2 do
12             if all arcs of ab_j have the same direction and are from a toward b then
13                 |ab_j^+| = |{(s, t) : (s, t) ∈ P, s appears before t on the path ab_j}|;
14             else
15                 |ab_j^+| = 0;
16         calculate max_i and orient C_i using the relation (3.2.1)
17         if max_i > max then
18             max = max_i;
19             C_max = C_i
20     return C_max
```

**Figure 9.** The algorithm for mixed cycles

## 3.3 Number of arcs and pairs

According to conjecture 3.1, the MMGO is probably not fixed-parameter tractable w.r.t. the number of pairs. On the other hand, it is clear that MMGO is not fixed-parameter tractable w.r.t. the number of arcs, $|A|$, otherwise, MTO could be solved in polynomial time. We show that MMGO is fixed-parameter tractable w.r.t. the combined parameter the number of pairs and the number of arcs. In the remainder of this section, let $|A| = k$.

Without loss of generality, we assume that the input graph $G$ is a mixed-acyclic graph [14]. One can see that the graph $G$ can be considered as a set of undirected trees which are connected using the arcs in $A$. We call each undirected tree an undirected component. For an illustration, consider the mixed graph in Figure 10, where the arcs are shown as dashed arrows.

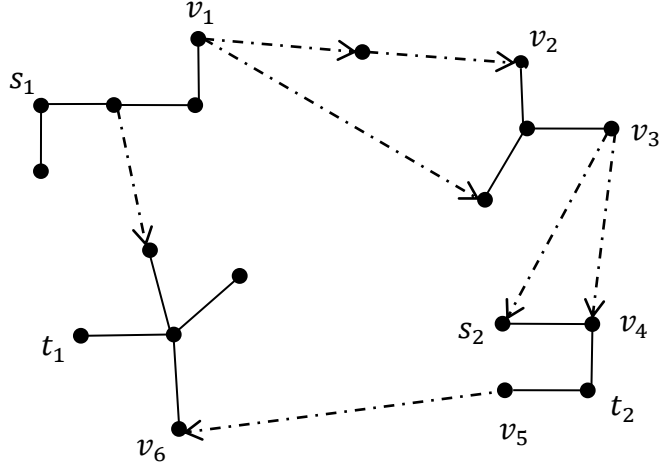

**Figure 10. Illustration of a mixed graph as connections of some undirected components**

The pairs in $P$ can be divided into two types $P_U$ and $P_D$. A pair $(s,t) \in P_U$ if both $s,t$ are in the same undirected component. Other pairs belong to $P_D$. Satisfying each type of pair can be handled as follows:

- Pair $(s,t) \in P_U$: in this case, there is only one path corresponding to the pair. Thus, this path must be oriented from $s$ to $t$ to satisfy the pair (such as $(s_2, t_2)$ in Figure 10).
- Pair $(s,t) \in P_D$: in this case, there may be more than one path corresponding to the pair. Thus, it is sufficient to orient one of them from $s$ to $t$ to satisfy the pair (such as $(s_1, t_1)$ in Figure 10). Each path may pass through one or more undirected components. If for each of this component, we determine the vertex through which a path enters the undirected component (the input vertex), and the vertex from which a path exits the undirected component (the output vertex), in fact, we have determined the path from $s$ to $t$, because in each undirected component there is only one path between the input vertex and the output vertex. Also, the remainder of the path is constituted by arcs.

Note that the input and output vertices are endpoints of the arcs that belong to an undirected component, thus, there are at most $2k$ of them. A vertex of an undirected component that is the head (tail) of an arc, is an input (output) vertex. For each vertex of an undirected component which is the endpoint of an arc, we must determine that the paths of which pairs pass through it. For example, the path corresponding to the pair $(s_1, t_1)$ may pass through the vertices $v_1, v_2, v_3, v_4, v_5, v_6$. Note that the vertex $s_i$ is considered as the input vertex through which the pair $(s_i, t_i)$ enters to the component containing $s_i$. Analogously, the vertex $t_i$ is considered as the output vertex from which the pair $(s_i, t_i)$ exits the component containing $t_i$.

For each consistent assignment of the pairs to the input and output vertices, some of the pairs in $P_D$ are satisfied. An assignment is consistent if for all components, the path from the input vertex to the output vertex, for all pairs that pass through that component, can be oriented. After orienting a consistent assignment, each undirected component has been converted to a mixed component. Since the pairs of $P_U$ in a component are independent of the pairs of $P_U$ in other components, the mixed components can be processed independently. According to proposition 3.1, the maximum number of pairs of $P_U$ that can be satisfied in each mixed component can be computed in $O(2^{|P_U|})$ time. The maximum number of total satisfied pairs among all consistent assignments of the pairs to the input and output vertices is the optimal solution to the problem. The algorithm is depicted in Figure 11.

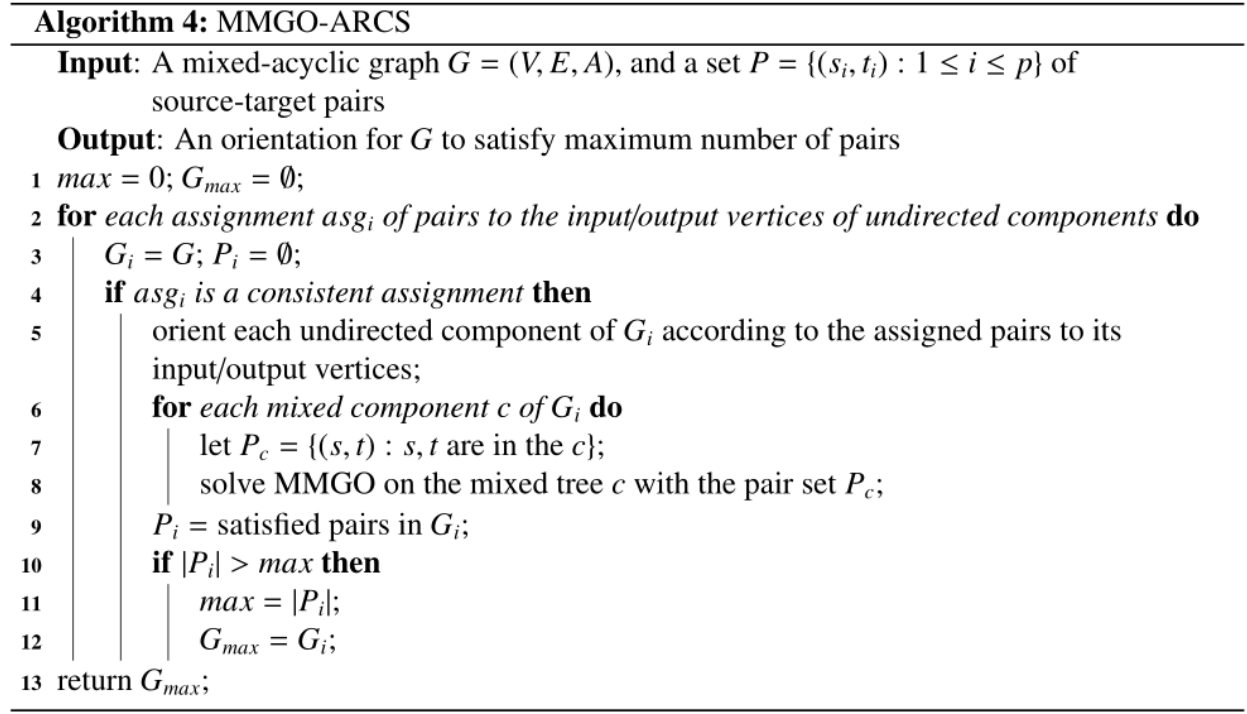

Figure 11. The algorithm for mixed graphs based on the number of pairs and arcs

**Theorem 3.3.1.** Given a mixed graph $G = (V, E, A)$ and a set $P = \{(s_i, t_i): 1 \leq i \leq p\}$ of source-target pairs, where $|A| = k$. The MMGO problem is fixed-parameter tractable w.r.t. the combined parameter $(p, k)$.

*Proof.* There are $O(2^{2pk})$ assignments of the pairs to the input and output vertices. Consistency checking of each assignment can be done in polynomial time. According to proposition 3.1, the MMGO problem can be solved on each mixed component in $O(2^p)$ time. Thus, the problem can be solved in $O(2^{(2k+1).p})$ time. Note that we have omitted some polynomial factors from the time complexity of the algorithm to simplify it. □

# 4 Conclusion

In this paper, we studied the maximum graph orientation problem on undirected and mixed graphs. In the undirected case, we determined the parameterized complexity of the problem (for non-fixed and fixed paths) w.r.t. the number of satisfied pairs, which was an open problem. Also, we presented an exact algorithm based on the number of leaves of a tree. In addition, we presented a parameterized-approximation algorithm w.r.t. a parameter named number of backbones of a tree. In the mixed case, we presented polynomial-time algorithms for paths and cycles, and an FPT-algorithm based on the combined parameter the number of arcs and the number of pairs for general graphs.

There are still some open problems for future investigations, some of which are:

- What is the parameterized complexity of the MMGO problem w.r.t. the parameter "number of pairs"?
- What is the parameterized complexity of the MTO problem w.r.t. the parameter "number of all pairs minus the number of input pairs"?
- Are there constant-ratio approximation algorithms for MTO and MMGO problems?